# Topological Semimetals for Scaled Back-End-Of-Line Interconnect Beyond Cu

Ching-Tzu Chen[1], Utkarsh Bajpai[1,2], Nicholas A. Lanzillo[3], Chuang-Han Hsu[4], Hsin Lin[4], and Gengchiau Liang[5]
[1]IBM Thomas J. Watson Research Center, Yorktown Heights, NY, USA, email: cchen3@us.ibm.com
[2]University of Delaware, Newark, DE, USA, [3]IBM Research, AI Hardware Center, Albany, NY, USA
[4]Institute of Physics, Academia Sinica, Taipei, Taiwan, [5]National University of Singapore, Singapore

*Abstract*—The resistance bottleneck in metal-interconnect scaling calls for new interconnect materials. This paper explores topological semimetals as a potential solution. After reviewing the desirable properties of topological semimetals for back-end-of-line (BEOL) interconnects, we use CoSi as an example to demonstrate the decreasing resistance-area product with scaling and provide material-search guidelines.

## I. Introduction

In today's integrated circuits, the RC delay of the BEOL Cu interconnect has become a major bottleneck of the system-level power-performance. When the metal linewidth scales below the electron mean-free path (39 nm for Cu), its resistivity increases in a power-law manner due to increased electron scattering off surfaces, grain boundaries and defects [1].

In contrast, a recent study shows that in NbAs, the resistivity (ρ) can *drop* by an order of magnitude from ρ = ~35 μΩ-cm in bulk single crystals to 1–5 μΩ-cm in ~200 nm nanobelts [2]. If such reduction persists below 10 nm, then NbAs interconnect lines would significantly outperform Cu. According to our benchmarking result, in the 5-nm-technology node with a 30-nm BEOL pitch, assuming ρ = 3 μΩ-cm with a 2-nm wetting layer yields a ~70% line-resistance reduction over Cu (Fig. 1a). This amounts to a substantial iso-power frequency uplift of >18% at 700W in circuit performance (Fig. 1b).

Why is the resistivity scaling in NbAs contrary to conventional metals? Briefly speaking, NbAs belongs to a new type of materials, called *topological semimetals* [3]. In an ordinary semimetal, the Fermi surface contains both electron and hole pockets. In a topological semimetal, the conduction and valence band form band-crossings at discrete nodal points (Figs. 2a, 2b) or along a loop (Fig. 2c) in the Brillouin zone (BZ) near the Fermi level. Furthermore, there exist non-trivial, disorder-tolerant *surface states* [4] passing through the band-crossing points (Fig. 2a), which may contribute the *same order of magnitude* to conduction as the bulk states [5].

This paper discusses the prospect of adopting topological semi-metals for BEOL interconnect applications, using two representative materials, CoSi and TaAs, as examples to illustrate the physics behind the enhanced current conduction with scaling, and concludes with a set of guiding principles for materials search.

## II. Topological Semimetals

The physics of topological semimetals is intimately related to that of topological insulators [3, 6]. One defining feature of topological semimetals is that the conduction and valence bands only touch at discrete points in the BZ. One can visualize a sphere enclosing each band-crossing node and define an integer topological invariant for the node as an integral associated with the Bloch wavefunctions over this fully gapped 2D surface. If the integral, called Chern number ($C$), is non-zero, then there must exist $|C|$ surface band(s) terminating at the so-called Weyl node in the projected surface BZ. Physics dictates that the net Chern number of all Weyl nodes in the BZ sums up to zero. Thus, Weyl nodes always appear in pairs with opposite chirality $\pm C$ (Fig. 2a, 2b), and the surface states connecting the pair of nodes form open *Fermi arcs* at the Fermi level, which is another defining feature of topological semimetals (Fig. 2a, Fig. 2d).

Many sub-classes of topological semimetals have been identified [7-10]. We roughly categorize them according to the dimensionality of the band crossings and band degeneracies at the nodes. Those with 0D band crossings and non-trivial Chern numbers include Weyl semimetals (Fig. 2a) and multifold-fermion semimetals (Fig. 2b). The former has a 2-fold band degeneracy, while the latter can have 3-, 4-, 6-, or 8-fold band degeneracies at the nodes. The best-known Weyl semimetals are the TaAs, TaP, NbAs, and NbP family [11, 3] with similar bandstructures, while the better studied multifold-fermion semimetals include CoSi, RhSi, and AlPt. [8-10, 12-14].

A well-separated pair of Weyl nodes cannot be removed unless the two nodes annihilate each other, so they are robust against weak perturbations. The presence of the Fermi arcs, guaranteed by the Weyl nodes, are thus also robust against small perturbations. In addition, in-depth analyses of the quasiparticle interference in TaAs show that intra-arc and inter-arc scatterings by surface adsorbents and surface defects are both strongly suppressed [15]. The small phase space further reduces the electron-phonon scattering of the Fermi-arc states, especially when the Fermi arc approaches a straight line [4]. These properties are highly desirable for scaled BEOL interconnect applications.

## III. Transport in nanoscale CoSi and TaAs films

CoSi is a promising candidate because its Fermi-arc states are well separated in energy from the bulk states except near the Weyl nodes (Fig. 2d). Furthermore, it is compatible with the Si-CMOS technology. Below we use CoSi as a model system to illustrate that (1) the Fermi-arc states indeed dominate the electrical conduction in nanometer-scale films and (2) Fermi-arc conduction is robust against surface defects.

Fig. 3 illustrates the resistance-area (RA) product of CoSi slabs with and without a surface notch to simulate strong surface disorder. In 8 – 40 atomic-layer (AL) CoSi slabs, $(RA)_{slab}$ is *significantly lower* than $(RA)_{bulk}$, in sharp contrast to Cu, and it continues to drop till ~2 nm. In 40AL slabs, the Fermi-arc contribution to total conduction exceeds $\geq$ 80% (Fig. 4b). This contribution increases gradually with decreasing thickness, consistent with the trend of decreasing RA.

The momentum-resolved transmission (Fig. 4a, 4d) shows that surface defects *cannot* eliminate the Fermi-arc conduction. Take the two states 1 and 2 in Fig. 4c and 5a as an example. Point 1 denotes a left-moving Fermi-arc state located on the top surface and point 2 denotes a right-moving state on the bottom. Fig. 4d shows that the transmission of the two states remains intact in the presence of surface defects because they are spatially separated. Instead of back-scattering, these electrons simply bend around the defects and continue forward (Fig. 5b).

Since the Chern numbers of the Weyl nodes at the BZ center and corner are $\pm 2$, there exist *2* robust connecting arcs (per surface, per spin), one of which extends from (0, 0) to ($\pi$, $\pi$) and the other from (0, 0) to (-$\pi$, -$\pi$). Additional surface channels are not protected from back-scattering. Neither are the bulk states. Therefore, the total conductance still reduces in the presence of defects. Fig. 6 shows that the percentage of the conductance loss due to surface defects decreases with increasing film thickness. In films above ~24 AL, the conductance loss stabilizes at ~25%, implying that 75% of the transmission is carried by the topologically protected Fermi-arc electrons.

What causes the excess conductance loss in thinner films? Fig. 7 and Fig. 8 show that, the closer the Fermi-arc state to Γ, the deeper it penetrates below the surface. Thus, in films below ~24 AL, more and more Fermi-arc electrons can back-scatter via the arc on the opposite surface. This puts a limit on the CoSi linewidth to ~2.5 nm for BEOL interconnect applications, consistent with the limit derived from the RA scaling in Fig. 3a. Our ab-initio quantum-transport simulation [17] predicts that $(RA)_{slab}$ can reduce to ~30% of $(RA)_{bulk}$ in the presence of defects. This hasn't accounted for the electron-phonon coupling. Since the electron-phonon scattering of the Fermi-arc electrons should be smaller than the bulk electrons [4], we expect that $(RA)_{slab}/(RA)_{bulk}$ would further decrease below 30%.

The corresponding RA scaling of (100)-terminated TaAs slabs (Fig. 9a) is summarized in Fig. 10a. Due to strong bulk-state interference, its $(RA)_{slab}/(RA)_{bulk}$ ratio is significantly higher than that of CoSi (Fig. 3, Fig. 10b). Besides, the surface density of states is anisotropic (Fig. 9b), so electrons must flow along [010] to maximize the surface-state contribution (Figs. 11). When spin-orbit coupling is small, the topological surface state connecting between the line nodes in bulk TaAs (Figs. 9b, 9c) provides a relatively robust surface-conduction mode (Fig. 12a). In nanoscale films, however, quantum confinement gives rise to extra surface modes (Fig. 12b). These electrons can backscatter at the surface defects, causing extra conductance loss *in addition to* the bulk-state loss. Only the transmission of the topological surface states remain intact.

## IV. CONCLUSION

The RA product of topological-semimetal thin films decreases with decreasing thickness, a highly desirable property for beyond-Cu interconnects. Their Fermi-arc surface states dominate the current conduction and remain robust against strong disorder. When the sample dimensions decrease below a threshold (~2.5 nm for CoSi), such a protection is weakened, revealing the scaling limit for BEOL interconnect applications. To further optimize the conductance scaling, we should search for materials with high Chern numbers, long and straight Fermi arcs spanning the entire BZ, and a large energy separation between the Fermi-arc surface states and the bulk states. To ensure manufacturability, other properties also need to be investigated, e.g., the adhesion to surrounding dielectrics, resistance to electromigration, and ability to withstand the 400 – 500C BEOL thermal budget.

ACKNOWLEDGMENT

The authors gratefully acknowledge the discussions with C. Lavoie, I. Garate, C.-Y. Huang, R. Sandararaman, S. Kumar, B. K. Nikolić, K. Dolui, T. M. Philip, B. Gotsmann, H. Schmid, K. Moselund, and J. J. Cha, and the funding support from the IBM Physical Sciences Council.

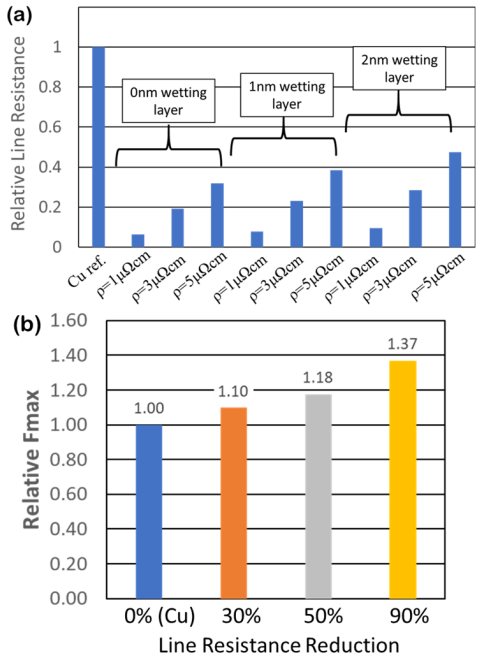

Fig. 1. Projected performance boost at 30 nm BEOL pitch, using sub-micron NbAs resistivity data in [2]: (a) relative line resistance and (b) circuit frequency uplift at 700W.

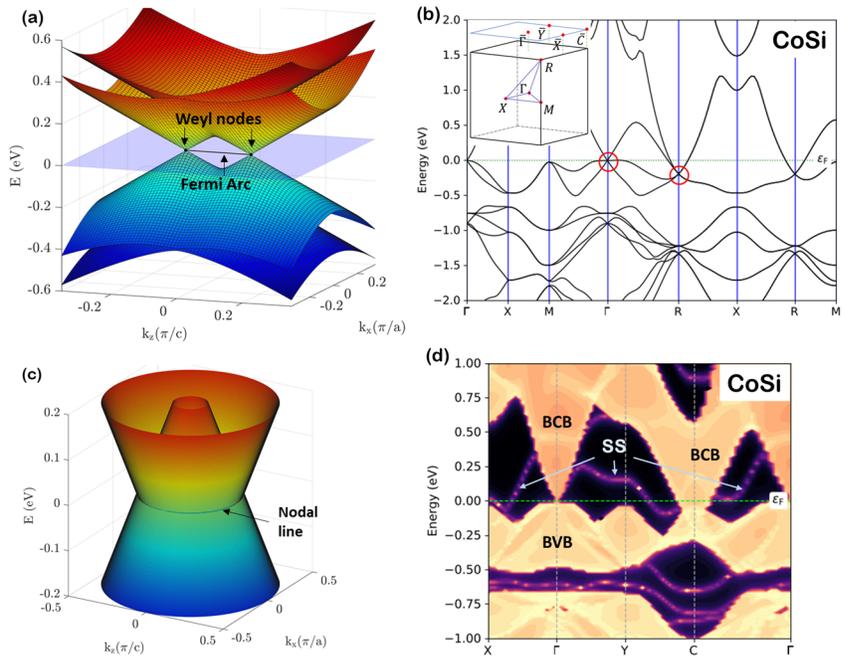

Fig. 2. Bandstructures of representative topological semimetals. (a) Model Weyl semimetal with one pair of Weyl nodes (with $C = \pm 1$) connected by a surface Fermi arc. (b) Prototypical multifold-fermion semimetal CoSi with $C = \pm 2$ at $\Gamma$ and R (marked in red circles) respectively. Inset: CoSi Brillouin zone. (c) Model nodal-line semimetal, where the band crossing forms a 1D ring. (d) Projected CoSi (100) surface spectral weight, showing that surface states (SS) connect between the projected $\Gamma$ and C points.

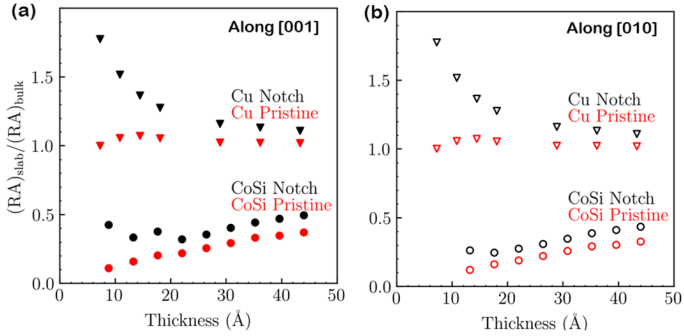

Fig, 3. Normalized resistance-area product (RA) of (100)-terminated CoSi slabs for transport along (a) [001] and (b) [010], compared against Cu.

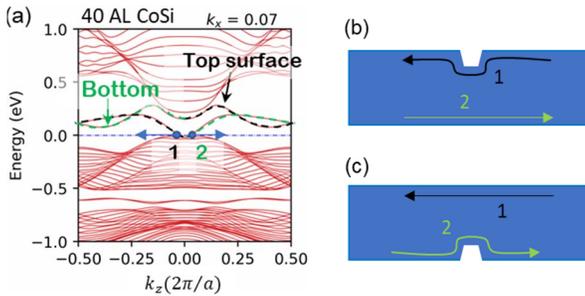

Fig, 5. (a) Bandstructure of a 40 atomic-layer (AL) (100)-terminated CoSi slab. Point 1 and 2 label the Fermi-arc states contributing to transport on the top and bottom surfaces, respectively. (Lattice constants of CoSi: a = b = c = 4.43 Å.) (b) Schematic of topological protection of the surface-state transmission in slabs with a notch on the top surface vs. (c) on the bottom surface.

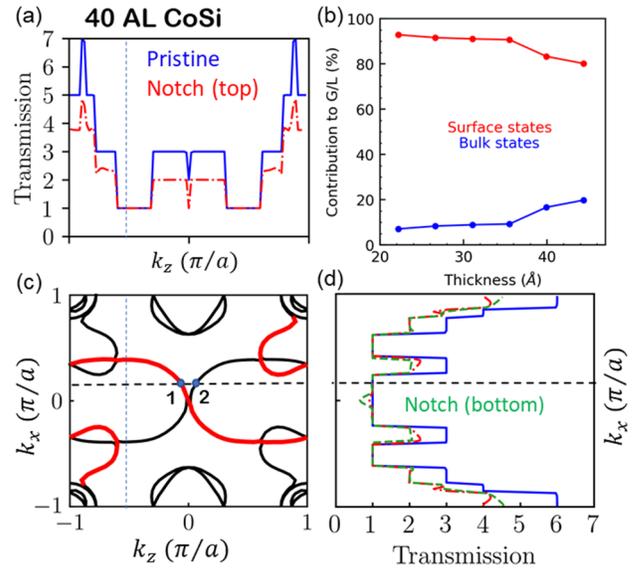

Fig. 4. Momentum $k$-resolved transmission of surface and bulk states in a 40-atomic-layer (AL) CoSi slab terminated at the (100) surface, with and without surface defects. (a) Transmission along the x-axis [010] and (d) z-axis [001] directions. (b) Surface vs. bulk contributions to the total conductance per unit length (G/L). (c) Fermi surface of a 40 AL slab, showing that Fermi-arc states emanate from the $\Gamma$ point to near the zone corner. Additional states at the corner are bulk quantum-well states. The Fermi-arc states labelled in red are located on the top surface. Their mirror symmetric counterparts across $k_z = 0$ are the Fermi-arc states on the bottom surface.

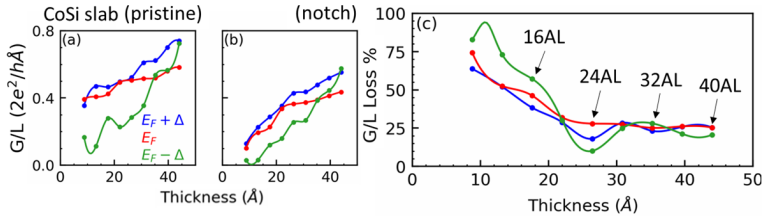

Fig. 6. Thickness dependence of the conductance per unit length (G/L) at the Fermi level and at 100 meV above and below the Fermi level in (a) ideal CoSi (100) slabs and (b) in slabs with a notch on the top surface. (c) Thickness dependence of the conductance loss due to scattering off the notch.

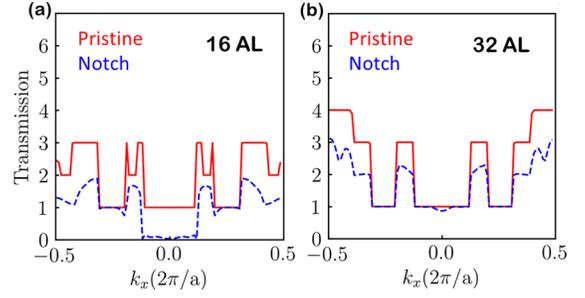

Fig. 7. Momentum-resolved transmission in the (a) 16AL and (b) 32AL CoSi slabs, illustrating the loss of ideal transmission of the Fermi-arc states near Γ due to top- and bottom-surface hybridization.

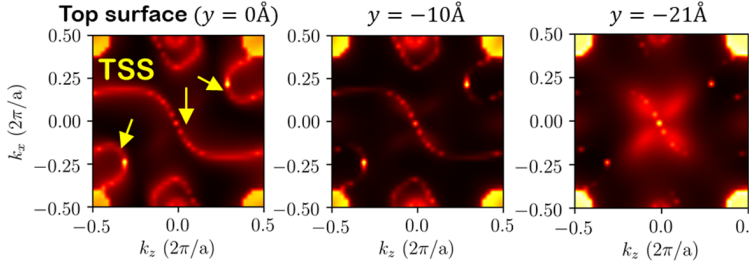

Fig. 8. Fourier-transformed local density of states at the CoSi (100)-surface and sub-surfaces, illustrating the $k$-dependent penetration depth of the Fermi-arc surface states (denoted by the yellow arrows). At 21Å below the surface, bulk states emerge around the Γ point, in addition to the deeply penetrating Fermi-arc states. (TSS stands for topological surface states.)

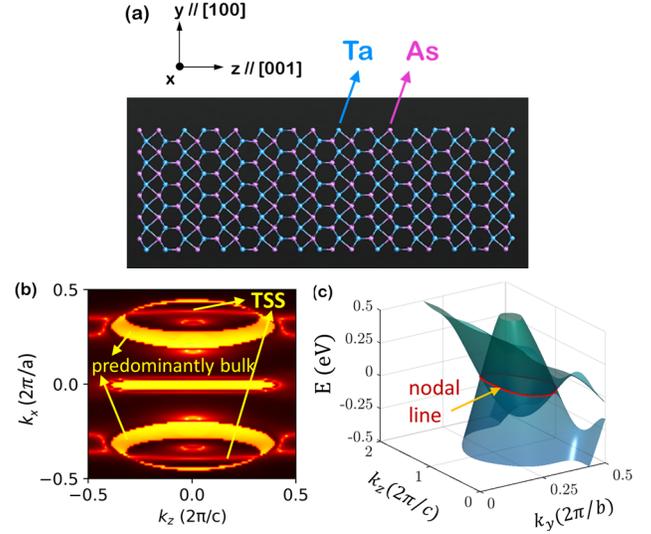

Fig. 9. (a) Schematic of a TaAs slab with the non-polar (100) surface termination. (TaAs lattice constants: a = b = 3.44 Å, c = 11.66 Å.) (b) Fourier-transformed local density of states of the TaAs (100) surface. (TSS: topological surface states.) (c) TaAs bandstructure at $k_x = 0$ in the absence of spin-orbit coupling, showing the mirror-symmetry protected nodal line.

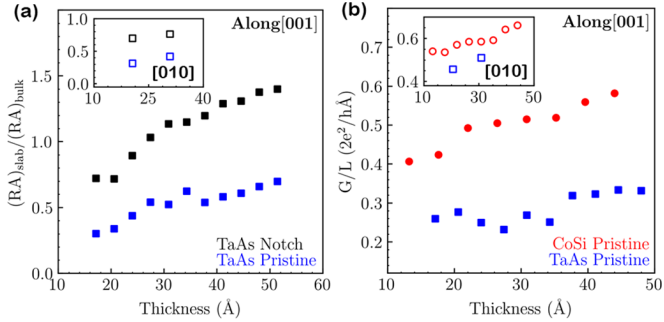

Fig. 10. (a) Resistance-area product of (100)-terminated ideal TaAs slabs and slabs with a notch on the top surface for transport along [001] (main panel) and [010] (inset). (b) Conductance per unit length (G/L) of pristine TaAs slabs, compared to CoSi, for transport along [001] (main panel) and [010] (inset).

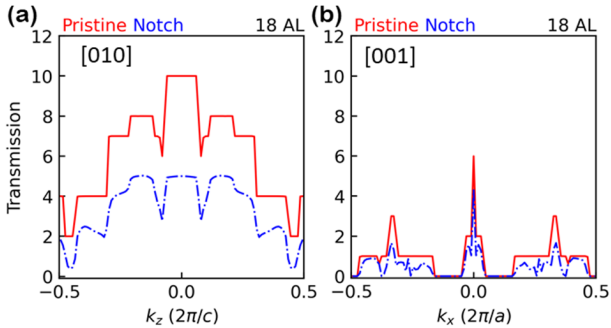

Fig, 11. Momentum-resolved transmission of surface and bulk states in an 18AL (100)-surface terminated TaAs slab along the (a) x-axis [010] and (b) z-axis [001] directions, with and without surface defects.

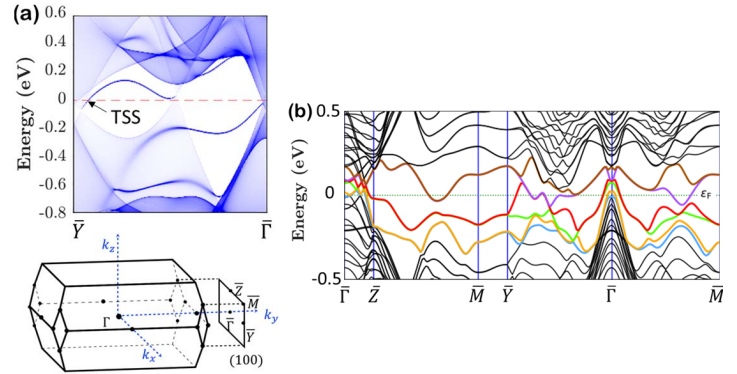

Fig. 12. (a) Top: surface spectral weight of (100)-surface-terminated semi-infinite TaAs bulk, where TSS denotes the topological surface states connecting between the nodal line (Fig. 9c). The TSS at Fermi energy provides a relatively robust surface mode for conduction. Bottom: TaAs Brillouin zone. (b) Bandstructure of a 30 AL TaAs slab, showing additional surface states at the Fermi level along $\bar{\Gamma} - \bar{Y}$ that are susceptible to defect scattering.